\documentclass[letter]{aa}         

\usepackage{natbib}
\usepackage{graphicx}
\usepackage{txfonts}
\usepackage{hyperref}

\begin{document}
\title{M Dwarf Exoplanet Surface  Density Distribution:} 
   \subtitle{ A Log-Normal Fit from 0.07-400 AU}

\author{Michael R. Meyer\inst{1,2} \and Adam Amara\inst{2} \and Maddalena Reggiani\inst{2,3} \and Sascha P. Quanz\inst{2}}

\institute{Department of Astronomy, University of Michigan, Ann Arbor, MI USA 48109 \\
\and Institute for Astronomy, ETH Zurich Wolfgang-Pauli-Strasse 27, Zurich, Switzerland, CH-8093\\
\and Space sciences, Technologies and Astrophysics Research (STAR) Institute, Universit\'{e} de Li\`ege, 19 All\'{e}e du Six Ao\^ut, B-4000 Li\`ege, Belgium}

\date{Received: June 4, 2016; Accepted July 14, 2017.}

\abstract{}{We fit a log-normal function to the M dwarf orbital surface density distribution of gas giant planets, over the mass range 1-10 times that of Jupiter, from 0.07-400 AU.}{We use a Markov Chain Monte Carlo approach to explore the likelihoods of various parameter values consistent with point estimates of the data given our assumed functional form.}{This fit is consistent with radial velocity, microlensing, and direct imaging observations, is well-motivated from theoretical and phenomenological viewpoints, and makes predictions of future surveys. We present probability distributions for each parameter as well as a Maximum Likelihood Estimate solution.}{We suggest this function makes more physical sense than other widely used functions, and explore the implications of our results on the design of future exoplanet surveys.}

\maketitle  
\titlerunning{M Dwarf Exoplanet Surface Density Distribution}
\authorrunning{Meyer et al.} 

\section{Introduction}


Confirmed exoplanet detections number well over a thousand enabling statistical studies of their orbital surface density and mass distributions, as well as overall frequency as a function of host star mass.  Radial velocity surveys (that probe the smallest orbital separations), as well as direct imaging surveys (that probe the largest separations) have been conducted for samples of M dwarfs 
\citep[0.1-0.6 M$_{\odot}$; e.g.][]{Bonfils2013,Bowler2015}, 
FGK stars \citep[0.5-1.5 M$_{\odot}$; e.g.][]{Mayor2011,chauvin2015}, and A stars \citep[1.5-2.5 M$_{\odot}$; e.g.][]{Johnson2010,Vigan2012}.  A pioneering attempt to constrain both the planet mass function and orbital surface density distribution for gas giant exoplanet populations surrounding FGK stars was made by \cite{Cumming2008} who fitted a homogeneous sample of RV detections, drawn from a stellar sample with well-known characteristics \citep{Fischer2005}.  These power-law fits revealed a rising planet mass function down to $>$ 0.3 times the mass of Jupiter (hereafter M$_{J}$) and a rising surface density distribution in units of logarithmic orbital period ($<$ 2000 days).  These fits have been used to plan, and interpret, a large number of direct imaging surveys representing hundreds of nights of 6-10 meter telescope time in the past decade \citep[e.g.][]{Lafreniere2007,Heinze2010,Macintosh2014,Beuzit2008}.  \footnote{Calibration of the evolutionary models needed to estimate masses from luminosities and temperatures is still underway representing an important caveat for direct imaging searches.}  Typical experiments assume a power-law surface density distribution rising with the log of the orbital radius \citep[cf.][]{Cumming2008} and then introduce an outer cut-off radius to truncate the population so that null results from direct imaging observations are consistent with the model \citep[e.g.][]{Reggiani2016,Nielsen2010}. 

Attempts to test whether similar fits can be used to describe exoplanet populations surrounding both higher and lower mass stars have revealed a dependence of either the overall planet occurrence, or the planet mass function on stellar mass: gas giant planets are detected with greater frequency around stars of higher mass \citep[e.g.][]{Johnson2010} at least up to 3.0 M$_{\odot}$ \citep{Reffert2015}.  The microlensing technique offers a unique opportunity to assess the exoplanet population at intermediate separations down to very low planet masses, predominantly around low mass M dwarf primaries \citep{Gould2010}.  However care must be taken in comparing the microlensing results to those from other techniques.  For one, we do not know for certain the host star mass of most of these microlensing events \citep{Fukui2015}.  Secondly, it is well known that bulk system metallicity impacts the outcome of gas giant planet formation \citep[e.g.][]{Valenti2005}:  the microlensing sample, probing the galactic bulge (3--10 Gyr), may be metal poor compared to younger field star samples probed by RV and direct imaging.  Knowing the orbital distribution of exoplanets, even over a limited mass range, is vital to test theories of planet formation and subsequent orbital evolution \citep[cf.][]{Mordasini2012}.

\begin{table*}[!h]
\caption{Observational Constraints on f (Number of 1-10 M$_{J}$ Planets Per Star) Versus Orbital Separation} 
\label{tab1}      
\centering                    
\begin{tabular}{lll}       
\hline\hline                 
Semi-major Axes (in AU) & f$_{Cassan^a}$ & Reference \\    
\hline                 
   0.07--0.33 & 0.008 $^{+0.0017}_{-0.005}$ & \cite{Bonfils2013}  \\      
   0.5-10.0  & 0.063 $\pm$ 0.0375 & \cite{Cassan2012} \\
   0-20  & 0.0625$\pm 0.03$ & \cite{Montet2014}  \\
   10-100 & $<$ 0.100 $\pm$ 0.001 (b) & \cite{Bowler2015}  \\
   8-400 & 0.022 $^{+0.028}_{-0.007}$ & \cite{Lannier2016}  \\ 
\hline                                  
\end{tabular}
\tablefoot{\\
\tablefoottext{a}{Assuming the planet mass function of \cite{Cassan2012}}\\
\tablefoottext{b}{For cold-start models, $<$0.154 $\pm$ 0.002}}\\
\end{table*}

Given the possibility of combining microlensing results with those from other techniques, low mass stars provide a unique opportunity to study exoplanet surface density distributions.  Previous attempts to reconcile surveys of gas giants surrounding M dwarfs with those of higher mass stars led to the conclusion that either the planet mass function surrounding M dwarfs must rise sharply to lower mass, or the normalization and/or orbital distribution must be different \citep{Quanz2012,Clanton2014}.  More recently \cite{Clanton2016} have shown that exoplanet statistics from RV \citep{Montet2014}, microlensing \citep{Gould2010}, and direct imaging \citep{Bowler2015} can be consistently fitted using a steep power-law in the surface density distribution that is abruptly truncated at orbital radii less than 15 AU.   Here we take an alternate approach and suggest a different, but well-motivated, functional form that also fits the most recent observational data.

\section{Why Log-Normal?}

It is well-known that the semi-major axis distribution of stellar binary companions surrounding FGK stars can be fit by a log-normal, peaking at about 50 AU \citep{Raghavan2010}.   The surface density of stellar companions surrounding M dwarfs and A stars also follow a log-normal with the mean proportional to the host star mass \cite{Janson2012,DeRosa2014}.  While models have been developed based on the idea that significant dynamical evolution in young star clusters is responsible for the observed log-normal period distribution \citep{Marks2012}, it is possible that the distribution is primordial, reflecting the outcome of the star formation process \citep{Parker2014,Offner2010}.  If stellar (and sub-stellar) companion formation surrounding stars (and any subsequent orbital evolution) results in a log-normal semi-major axis distribution, perhaps planet formation does too.   While gravitational instability probably only occurs in rare circumstances \citep{Vigan2017}, it could give rise to different mass and orbital separation distributions compared to the core accretion model \citep{Santos2017}.  We know that gas planet formation via core accretion is complex, resulting from many independent processes 
(growth of solids, build up of critical mass core, runaway gas accretion) and further evolution of their orbital radii depends on many factors \citep{Benz2014,Helled2014}.  In the limit of a product of an infinite number of independent variables, regardless of their underlying distributions, the central limit theorem dictates an outcome in the form of a log-normal.  In reality a limited number of variables rapidly approaches a log-normal \citep{Adams1996}).  Thus far, the data for stars over a wide mass range suggest that exoplanet populations exhibit a log-rising frequency at small separations \citep[< 3 AU;][]{Cumming2008} but are not common at orbital radii > 30 AU \citep{Bowler2016}.  It seems to us that a log-normal distribution is a reasonable choice of functional form to describe the semi-major axis distribution of exoplanets that appears to rise with logarithmic separation, and then fall.  Here we demonstrate that the data for M dwarfs are consistent with such a distribution.

\section{Observational constraints on orbital surface density distribution}

Key to our approach is the explicit assumption that the planet mass function does not depend on orbital separation:  we are not aware of any data that would require us to reject this null hypothesis  (excluding the pileup of hot Jupiters within 0.1 AU).  We also adopt a planet mass function consistent with available data in order to inter-compare various surveys.  We have adopted constraints on the frequency of exoplanets around M dwarf primaries over a common range of planetary mass from 1-10 M$_{J}$ and diverse orbital separations from a number of different surveys.    These results are summarized in Table\ref{tab1}.  From the survey of \cite{Bonfils2013} we consider a range of separations from 0.07 to 0.33 AU.  Bonfils et al. quote a frequency of $0.02 ^{+ 0.03}_{-0.01}$ for the average number of planets between 0.3-3 M$_{J}$ per star based on their data.  We have adjusted their result to our common mass range assuming the mass function of \cite{Cassan2012} to ensure consistency with the microlensing results discussed below.  We also checked that in the range of masses where the survey is complete, the distribution of masses implied by RV observations were consistent with this assumed mass function (based on the available inclination estimates, or assuming the expectation value of 60$^\circ$ for the inclination).  The primary stars in the \cite{Bonfils2013} survey range from 0.1--0.6 M$_{\odot}$.  These results are consistent with previous other surveys \citep[e.g.][]{Endl2006} and can also be reconciled with M dwarf results presented in \cite{Cumming2008} given the uncertainties.

At intermediate separations, we consider results from microlensing surveys that are particularly sensitive to planets in the separation range 0.5-10 AU assuming all hosts are M dwarfs.  Adopting the results of \cite{Cassan2012} for the frequency as well as the planet mass function, we estimate the frequency of planets between 1-10 M$_{J}$ in the above separation range to be 0.063 $\pm 0.0375$ planets per star.  Although these detections might also include the detection of very low mass brown dwarf companions, such objects are expected to be relatively uncommon given the universal companion mass ratio distribution of \cite{Reggiani2013}.  This expectation is confirmed in the recent analysis by \cite{Shvartzvald2016} where the companion mass ratio distribution for microlensing events (assuming M dwarf primaries) shows a local minimum above 10 M$_J$, consistent with the model for FGK stars in \cite{Reggiani2016}.

For the largest orbital separations, we utilize results from direct imaging surveys.  Here we quote results from \cite{Montet2014}, \cite{Bowler2015}, and \cite{Lannier2016} regarding the frequency of gas giant planets surrounding primaries between 0.2--0.6 M$_{\odot}$.  \cite{Montet2014} combine results from a long time  baseline radial velocity survey of 111 M dwarfs with direct imaging constraints that inform the interpretation of accelerations observed without full orbital solutions.  They report a frequency of 0.065$\pm 0.03$ planets per star over the mass range 1--13 M$_{J}$ between 0-20 AU.  \cite{Bowler2015} quote an upper limit of 0.103 planets per star over the range from 10-100 AU and 1-13 M$_J$ based on hot--start models of their early evolution:  if cold--start models are considered this limit is recalculated to be 0.16.    Most recently, \cite{Lannier2016} report a high contrast imaging survey of 54  M dwarfs resulting in an estimate for the frequency of 0.023 $^{+0.029}_{-0.007}$ over a mass range 2--14 M$_{J}$ between 8--400 AU.  We adjusted these results assuming a planet mass function from \cite{Cassan2012} for comparison to other surveys over a common mass range of 1-10 M$_{J}$ as presented in Table~\ref{tab1}.  

\section{Results}

   \begin{figure}

   \centering
   \includegraphics[width=8cm]{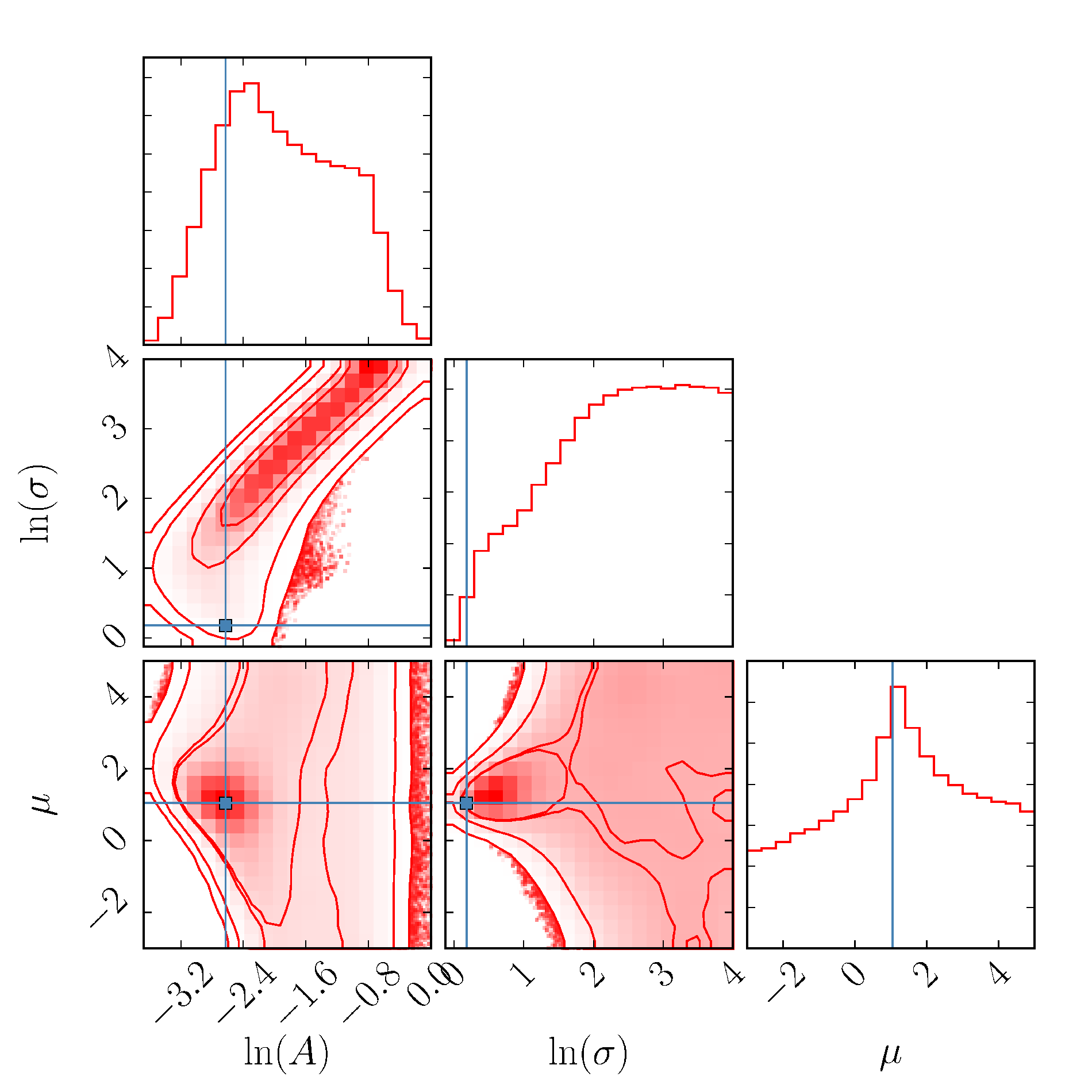}
      \caption{
Probability density functions for the three parameters used in the log--normal fit described in the text:  ln(A), ln($\sigma$), and $\mu$ (red histograms). Also shown are the correlations between parameters where reds are highest and whites are lowest values of the likelihoods. Also noted in blue are the locations of the parameter values in the MLE fit described in the text.}
   \label{fig1}
   \end{figure}

Using constraints from these data (including the upper limit), and our assumption of a log-normal functional form, we explore allowed values of three parameters (amplitude ($A$), mean of the log-normal ($\mu$) , and square root of the variance ($\sigma$), all restricted to be positive) using a Markov Chain Monte Carlo approach to survey the landscape of the likelihood function with the publicly available package CosmoHammer \citep{Akeret2013}.  We adopted a probability density function (hereafter PDF) for the RV planet frequency from \cite{Bonfils2013} \citep[as well as the direct imaging results of][]{Lannier2016} that accurately reflect the reported asymmetric error bars (a log-normal).  For the microlensing results \citep{Cassan2012} as well as \citep{Montet2014} we assumed a gaussian with a mean and sigma corresponding to the reported values and uncertainties.  We adopted a complementary error function for the \cite{Bowler2015} hot--start results with the mean as the upper limit and a sigma of 0.001.  We assume flat priors in the log for each free parameter, with $-4 < ln(\sigma) < 4$ \footnote{Beyond $ln(\sigma) > 2$ all fits are equal, and not as good as fits with $ln(\sigma) < 2$ but are difficult to formally rule out, which is why we quote the maximum likelihood estimate below.}.  We then obtain a PDF for each variable, marginalized over the other two variables, as well as the correlations between variables as shown in Figure~\ref{fig1}.  While the amplitude and mean are reasonably well constrained, the width of the distribution is not.  This results in the degeneracy between $\sigma$ and amplitude as well as structure in the $\sigma$ versus $\mu$ plot: when $\sigma$ becomes very large, $\mu$ is poorly constrained and the amplitude is adjusted accordingly.  We also calculate the Maximum Likelihood Estimate (MLE), that maximizes the chance of these data being drawn from our model, the parameter values of which are also denoted in Figure~\ref{fig1}.   Where $\phi(x)$ is the PDF of having a gas giant planet 1-10 M$_J$ as a function of the orbital semi-major axis x and f is the frequency over the limits of integration:

\begin{equation}
\phi(x) = \frac{df}{dx} = A \frac{e^{(ln(x)-\mu)^2/2\sigma^2}}{x 2 \pi \sigma}
\end{equation}

\begin{figure}
   \centering
   \includegraphics[width=8cm]{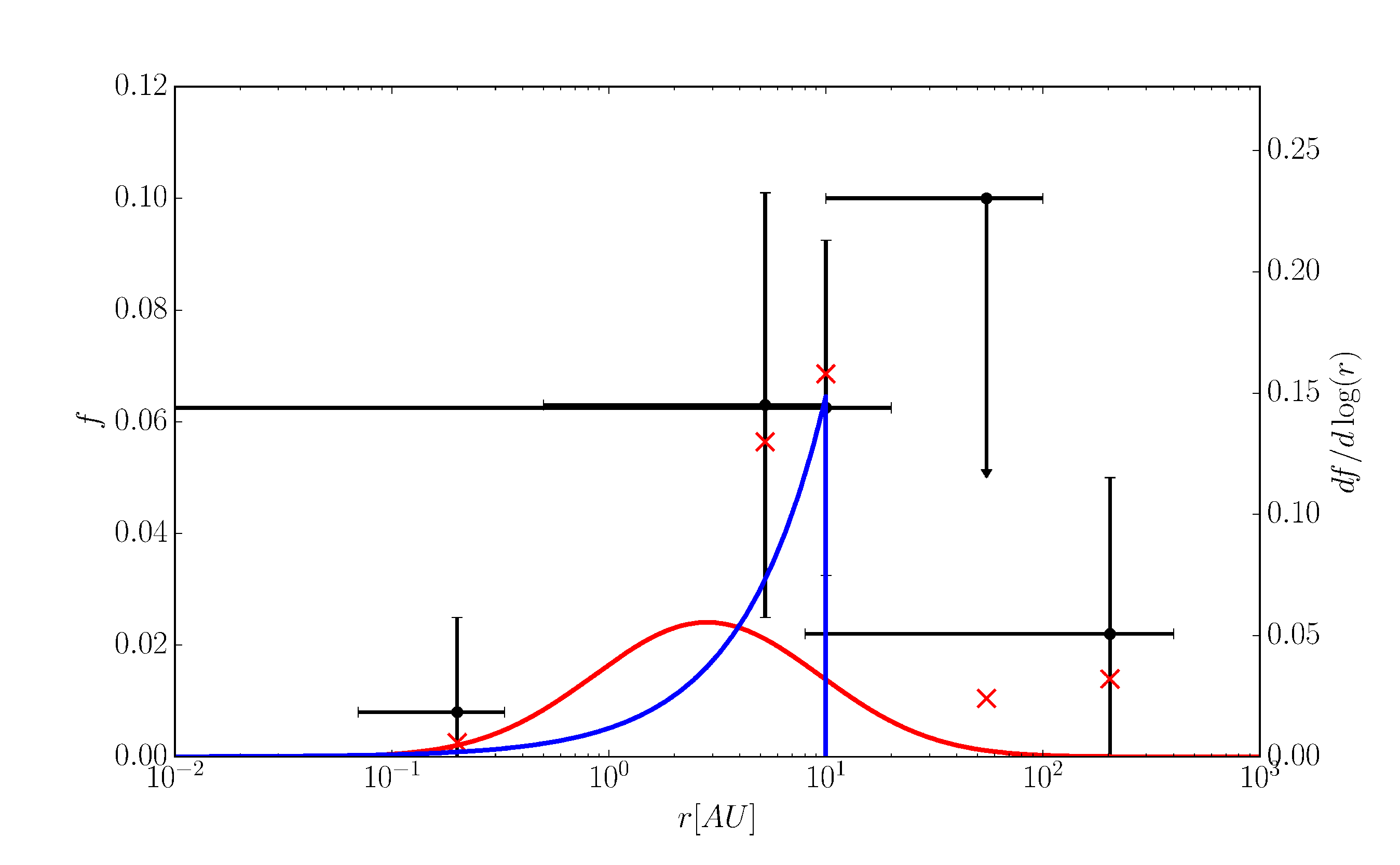}
      \caption{
Log-normal surface density distribution of 1-10 M$_{J}$ Companions to M dwarfs.  On the left axis are the integrated values of f, where the black points correspond to the data points listed in Table~\ref{tab1}: note the horizontal lines are ranges while the vertical lines are errors in f.  On the right axis are the relative values of the differential PDFs.  In red is the MLE fit as described in the text.  The integrated f values for the MLE, over the ranges in semi-major axis corresponding to the data from Table~\ref{tab1}, are shown as red ''x''.  Also shown in blue is the fit from \cite{Clanton2016} which was published before the \cite{Lannier2016} results were available. 
}
   \label{fig2}
   \end{figure}

The MLE has parameters $ln(A) = -2.6267$, $ln(\sigma)=0.1801$, $\mu = 1.0413$.  \footnote{In base ten $\phi(x) = A \frac{e^{(log(x)-\mu)^2/2\sigma^2}}{2 \pi \sigma}$ with same amplitude A, $\mu = 0.4527$, and $\sigma=0.5205$).} 
We present this fit (in red), in Figure ~\ref{fig2}, as well as the integrated frequencies over the ranges indicated from our fits for comparison with the data in Table~\ref{tab1}.  This distribution, with a mode at 2.8 AU yields 0.07 planets per star over the mass range from 1-10 M$_J$ over all separations \citep[assuming a planet mass function from][]{Cassan2012} and satisfies the observed constraints.  

\section{Discussion}

In the MLE, half of all planets with masses 1--10 M$_J$ are within 2.8 AU and half are beyond with a prediction of $0.015 < f < 0.03$ (at 80 \% confidence) from 10-1000 AU.  Recently \cite{Clanton2016} fitted similar data (excluding the new \cite{Lannier2016} result) with a continuous power-law that rose sharply to account for the microlensing detections, but then truncate abruptly the planet population due to the direct imaging upper limits from \cite{Bowler2015} (shown in blue in Figure~\ref{fig2}), implying no gas giant planets beyond 10 AU.  We prefer the log-normal to model this rise and fall of the orbital surface density distribution of gas giant exoplanets, which is also consistent with the new  \cite{Lannier2016} estimates indicating a small, but non--zero fraction of gas giant planets beyond 10 AU. 

Additional observations that would better constrain the width of the log-normal are crucial.  For example, next generation microlensing surveys could significantly improve upon the precision of the estimated frequency of exoplanets between 0.5--10 AU \citep[e.g.][]{Gaudi2012}.  Future high contrast imaging surveys with ground- and space-based telescopes will help place important constraints on $\sigma$.  For example, a survey with the NIRCam instrument on JWST will be able to measure $f$ much more precisely, down to much lower masses, from 10-100 AU around very nearby, young, M dwarfs \citep[e.g.][]{Schlieder2016}.  In the next decade ELTs will be able to image planets across this expected peak in the log-surface density distribution at about 3 AU.  Should the location of the peak in the log-normal surface density distribution depend on stellar mass?  If related to photoevaporation (as a stopping point in the inner migration of gas giant planets formed at large separations) perhaps yes \citep[e.g.][] {Alexander2012,Ercolano2015} in that lower mass stars should have peaks closer to their stars.  However, phenomenological predictions of core accretion, relating the time to reach a critical core mass for runaway gas accretion as a competition between disk lifetime (which depends inversely on stellar mass) and the rate of collisions (faster around higher mass stars with more massive disks) suggest a weak dependence.   On the other hand, if the water ice line is critical to gas giant planet formation, one expects a linear dependence of this quantity on stellar mass in the pre-main sequence where $L \sim M^2$ \citep[c.f.][]{Kennedy2008}.  Assuming that the overall frequency of gas giant planet formation depends linearly on stellar mass, our model predicts $f = 0.06$ for the frequency of  planets 1-10 M$_J$ between 10-100 AU around solar mass stars, or even higher if the mode of the log-normal depends linearly on stellar mass.  So far, these numbers are consistent with estimates for the frequency of gas giants on wide orbits from legacy RV surveys \citep[see][]{Wittenmyer2016,Durkan2016} but will be put to a strong test by on--going FGK star surveys such as those being carried out now with SPHERE and GPI.  Extrapolating the amplitude of our fitted function to A stars with a linear mass dependence, but not adjusting the mode, we could expect f = 0.12 from 1-10 M$_J$ from 10-100 AU surrounding A star samples, consistent with current constraints \citep[e.g.][]{Vigan2012}.  In a future paper we will explore concretely the stellar mass dependence of exoplanet populations with a focus on the planet mass function. 

\begin{acknowledgements}
We thank an anonymous referee for a helpful review, as well as other anonymous referees who commented on earlier versions of this manuscript.  This work has been carried out in part within the frame of the National Centre for Competence in Research PlanetS supported by the Swiss National Science Foundation. MRM and SPQ are pleased to acknowledge this financial support of the SNSF.   M.R. acknowledges funding from the European Research Council Under the European Union’s Seventh Framework Program (ERC Grant Agreement No. 337569) and from the French Community of Belgium through an ARC grant for Concerted Research Action.
\end{acknowledgements}

\bibliographystyle{aa} 
\bibliography{lognormal} 

%

\end{document}